\documentclass[12pt,a4p]{article}
\usepackage{epsfig}
\usepackage{amsmath,times}

\def\pin{$\rm \pi^0\ $}
\def\pins{$\rm \pi^0s\ $}

\def\Zn{$\rm Z^0\ $}
\newcommand{\etal}      {{\it et~al.}}

\newlength{\picwi}
 \newcommand{\tcaption}[1]{
        \refstepcounter{table}
        \setbox\@tempboxa = \hbox{\footnotesize \bf Table~\thetable. #1}
        \ifdim \wd\@tempboxa > 6in
           {\begin{center}
        \parbox{6in}{\footnotesize\baselineskip=12pt \bf Table~\thetable. #1}
            \end{center}}
        \else
             {\begin{center}
             {\footnotesize \bf Table~\thetable. #1}
              \end{center}}
        \fi}
\begin{document}
\begin{titlepage}
\begin{center}{\large   EUROPEAN ORGANIZATION FOR NUCLEAR RESEARCH
}\end{center}\bigskip
\begin{flushright}
       CERN-EP/2003-005   \\ 28th January 2003
\end{flushright}
\bigskip\bigskip\bigskip\bigskip\bigskip
\begin{center}{\huge\bf   
        Bose-Einstein Correlations of \pin Pairs\\
        from Hadronic $\rm \bf{Z^0}$ Decays
}\end{center}\bigskip\bigskip
\begin{center}{\LARGE The OPAL Collaboration
}\end{center}\bigskip\bigskip
\bigskip\begin{center}{\large  Abstract}\end{center}
   We observe Bose-Einstein correlations in \pin pairs produced
   in \Zn had\-ronic decays using the data sample 
   collected by the OPAL detector at LEP 1
   from 1991 to 1995.  
   Using a static Gaussian picture for the 
   pion emitter source,
   we obtain the chaoticity parameter    
   $ \lambda = 0.55 \pm 0.10 \pm 0.10 $ and the source radius  
   $   R = (0.59 \pm 0.08 \pm 0.05)\ \rm fm. $
   According to the JETSET and HERWIG
   Monte Carlo models, the Bose-Einstein correlations in our data
   sample largely  
   connect \pins originating from the
   decays of different hadrons. Prompt pions formed at string 
   break-ups
   or cluster decays only form a small fraction of the sample.
\bigskip\bigskip\bigskip\bigskip
\bigskip\bigskip
\begin{center}{\large
(Submitted to Physics Letters B)
}\end{center}
\end{titlepage}
\begin{center}{\Large        The OPAL Collaboration
}\end{center}\bigskip
\begin{center}{
G.\thinspace Abbiendi$^{  2}$,
C.\thinspace Ainsley$^{  5}$,
P.F.\thinspace {\AA}kesson$^{  3}$,
G.\thinspace Alexander$^{ 22}$,
J.\thinspace Allison$^{ 16}$,
P.\thinspace Amaral$^{  9}$, 
G.\thinspace Anagnostou$^{  1}$,
K.J.\thinspace Anderson$^{  9}$,
S.\thinspace Arcelli$^{  2}$,
S.\thinspace Asai$^{ 23}$
% ...
}\end{center}\bigskip
\bigskip
$^{  1}$School of Physics and Astronomy, University of Birmingham,
Birmingham B15 2TT, UK
\newline
$^{  2}$Dipartimento di Fisica dell' Universit\`a di Bologna and INFN,
I-40126 Bologna, Italy
\newline
$^{  3}$Physikalisches Institut, Universit\"at Bonn,
D-53115 Bonn, Germany
\newline
% ...
$^{  5}$Cavendish Laboratory, Cambridge CB3 0HE, UK
\newline
% ...
$^{  9}$Enrico Fermi Institute and Department of Physics,
University of Chicago, Chicago IL 60637, USA
\newline
% ...
$^{ 16}$Department of Physics, Schuster Laboratory, The University,
Manchester M13 9PL, UK
\newline
% ...
$^{ 22}$Department of Physics and Astronomy, Tel Aviv University,
Tel Aviv 69978, Israel
\newline
$^{ 23}$International Centre for Elementary Particle Physics and
Department of Physics, University of Tokyo, Tokyo 113-0033, and
Kobe University, Kobe 657-8501, Japan
\newline
% ...

\bigskip
\par
\section{Introduction }  
  The Bose-Einstein correlations (BEC) effect has a quantum-mechanical origin.
  It arises from the requirement to 
  symmetrise the wave function of a system of two or more 
  identical bosons.
 It was introduced 
 into particle reactions leading to multi-hadron final states
 as the GGLP effect~\cite{GGLP} in the study of the 
 $\pi^+\pi^+$ and $\pi^-\pi^-$ systems. 
 The distributions of the 
 opening angle between the momenta in pairs of like-sign pions were
 shifted towards smaller values compared to the corresponding distributions 
 for unlike-sign pairs. A related effect was exploited earlier 
 in astronomy~\cite{HBT} to measure the radii of stars.\par 
 In high energy physics, for example $\rm e^+e^- $ collisions at LEP, 
 a quantitative understanding of the BEC effect allows tests of  
 the parton fragmentation and hadronisation models. This would in turn 
 help in achieving a more precise measurement of  
 the W boson mass and better knowledge of several
 Standard Model (SM) observables~\cite{SM}.
 The fragmentation models presently used are those of strings and clusters 
 implemented, respectively,  in the JETSET~\cite{jetset}
 and HERWIG~\cite{herwig} Monte Carlo generators.\par
 Numerous studies of BEC in pairs of identical bosons already exist, see for
 example~\cite{opal0}.
 Due to the experimental difficulties in photon and
\pin reconstruction, only very few studies~\cite{l3paper}
 exist for BEC in   
\pin pairs, even though they offer  
 the advantage of being free of final state Coulomb corrections.\par
 The string model
 predicts a larger BEC strength or chaoticity  
 and a smaller effective radius of the emitting source 
 for \pin  pairs compared to 
 $ \pi^{\pm} $ pairs
 while the cluster fragmentation model predicts the same
 source strength and size~\cite{anderson, zalewski}. However,   
 neither model of primary hadron production  
 has a mechanism to allow BEC between \pins produced in 
 different strong decays. 
 The string model prediction is
 a consequence of electric charge conservation 
 in the local area where the string breaks up.
 Similar expectations can be 
 derived if the probabilities in the string break-up mechanism are interpreted
 as the squares of quantum mechanical amplitudes~\cite{anderson,others}. 
 A small difference between 
 $ \pi^{\pm} $ pairs and \pin pairs is also expected from a pure 
 quantum statistical approach to Bose-Einstein symmetry~\cite{andreev}. 
 In addition,  
 based on isospin invariance, suggestions exist on 
 how to relate BEC in the pion-pair systems  
 i.e.\ $\pi^0\pi^0,$ $\pi^{\pm}\pi^{\pm},$
 and $\pi^+\pi^-$ and how to extend it to $\pi^{\pm}\pi^0$.~\cite{Lipkin}. 
 The L3 collaboration has recently reported~\cite{l3paper}
 that the radius of the neutral-pion source may be smaller 
 than that of charged pions,
$ R_{\pi^{\pm}\pi^{\pm}}-R_{\pi^0\pi^0}=(0.150\pm0.075 
 \rm (stat.)\pm 0.068(syst.))
 \rm \ fm,$
 in qualitative agreement
 with the string fragmentation prediction.\par 
 This paper presents a study of BEC in \pin pairs  using
 the full hadronic event sample 
 collected at centre-of-mass energies at and near the $\rm Z^0 $ peak by the 
 OPAL detector at LEP from 1991 to 1995.  
 This corresponds to about four million 
 hadronic $\rm Z^0 $ decays. A highly pure sample of \pin mesons is  
 reconstructed 
 using the lead-glass electromagnetic calorimeter.
 The correlation function is obtained after accounting for purity and 
 resonant background. It
 is parametrised with a static picture of a Gaussian emitting source~\cite{GGLP,HBT}.\par
\section{Selection of hadronic $\rm \mathbf{Z^0} $ decays}
 A full description of the OPAL detector can be found in~\cite{opald}. 
 The sub-detectors relevant to the present analysis are the 
 central tracking detector
 and the electromagnetic calorimeter.
 The central tracking detector consists of a silicon micro-vertex
 detector, close to the beam pipe, and three drift chamber devices:
 the vertex detector, a large jet chamber and  surrounding
 $z$-chambers\footnote{The OPAL coordinate system 
      is defined so that the $z$ axis is in the
      direction of the electron beam, the $x$ axis 
      points towards the centre of the LEP ring, and  
       $\theta$ and $\phi$
       are the polar and azimuthal angles, defined relative to the
      $+z$- and $+x$-axes, respectively. 
      In cylindrical polar coordinates, the radial coordinate is denoted
      $r$.}.
 In combination, the three drift chambers 
 sitting inside a solenoidal magnetic field of 0.435 T
 yield a momentum resolution of
 $\sigma_{p_t}/p_t \approx \sqrt{0.02^2+(0.0015\cdot p_t)^2}$
 for $|\cos(\theta)| < 0.7$, where $p_t$ (in GeV) is the transverse momentum
 with respect to the beam axis.
 The electromagnetic calorimeter detects and measures the energies and 
 positions of electrons, positrons and photons 
 for energies above 0.1 GeV.
 It is a total absorbing calorimeter, and is 
 mounted between the coil and the iron yoke of the magnet.
 It consists of 11704 lead-glass blocks arranged in
 three large assemblies (the barrel that surrounds 
 the magnet coil, and two endcaps)
 which together cover 98\% of the solid angle. 
   The intrinsic energy  resolution is 
    $ \sigma_E/E\simeq 5\% / \sqrt{E}, $ where $ E $  is  
   the electromagnetic energy in GeV.\par
   Standard OPAL selection criteria are applied to tracks and
   electro\-magnetic clus\-ters~\cite{billg}. 
   Tracks are required to have at least
   20 measured points in the jet chamber, a measured
   momentum greater than $\rm 0.1\ GeV $, an impact 
   parameter $|d_0|$
   in the $r-\phi$ plane smaller than 2~cm, a $z$ position 
   at the point of closest approach to the origin in the
   $r-\phi$ plane within 
   25 cm of the interaction point, and a measured polar angle with 
   respect to the beam axis
   greater than $ \rm 20^\circ .$
   Electromagnetic
   clusters are required to have an energy greater 
   than $\rm  0.1\ GeV $ if they 
   are in the barrel part of the detector (i.e.\  
   $\mid\cos{\theta}\mid\ \le 0.82 $) 
   or greater than $\rm 0.3\ GeV $ if they are in the endcap parts. 
   Hadronic $\rm Z^0 $ decays are selected by 
   requiring for each event more than 7 measured tracks, a visible 
   energy 
   larger than 60 $\rm GeV $ and an angle
   larger than $\rm 25^\circ$ and smaller than $\rm 155^\circ$
   between the calculated event 
   thrust~\cite{thrust} axis and the beam axis.
   The visible energy is the energy sum of all detected tracks,   
   electromagnetic clusters not associated to tracks and 
   electromagnetic clusters associated to tracks after correcting
   for double counting.  
   A sample of 3.1  million $\rm Z^0 $ hadronic decays is selected for which  
   the total background, consisting mainly of $\tau $ pairs, 
   is less than $\rm 1 \% $ and is neglected
   throughout the analysis. 
\par
 Detector effects and detection efficiencies for the  
 spectra of \pin pairs are evaluated
 using eight million Monte Carlo hadronic
 $\rm Z^0 $ decays. Events are generated using the JETSET 7.4 program, 
 tuned to reproduce the global features of hadronic events as measured 
 with the OPAL detector~\cite{billg}, with the BEC effect explicitly 
 switched off. Samples generated with the HERWIG 5.9 
 program without the BEC effect are used for comparison.
 The generated events were passed through a full simulation of the 
 OPAL detector~\cite{opalsim} and were analysed using the same 
 reconstruction and selection programs as were applied to the data.\par
\section{ Reconstruction of $ \pi^0 $ mesons }  
  For the selected event sample, 
  neutral pions are reconstructed from photon pairs.
  Photon reconstruction is performed in the barrel part of the 
  electromagnetic calorimeter  
  where both the photon reconstruction efficiency and the energy 
  resolution are good. 
  The procedure of~\cite{opal1}  which resolves photon candidates 
  in measured electromagnetic clusters is used. It employs
  a parametrisation of the expected lateral energy 
  distribution of electromagnetic showers. It is optimised to 
  resolve as many photon candidates as possible from the 
  overlapping energy deposits in the electromagnetic calorimeter in a
  dense environment of hadronic jets.  The purity of the photon 
  candidate sample is further increased using  a
  likelihood-type function~\cite{opal1} 
  that associates to each photon candidate a 
  weight $ w $ for being a true photon.
  Photon candidates with higher $w$ are more likely to be true photons.\par
  All possible pairs of photon candidates are then 
  considered. Each pair was assigned a probability $ P $ for both 
  candidates being correctly reconstructed as photons. 
  This probability is simply the
  product of the $w$~-weights  associated with the two candidates.
  The combinatorial background consists of a mixture of three
  components: (i) wrong pairing of two 
  correctly reconstructed  photons, (ii) pairing of two fake photons 
  and (iii) pairing of one correctly reconstructed photon with 
  a fake one. Choosing only photon pairs with high
  values of $ P $ leaves combinatorial background 
  mostly from component (i).\par
\begin{figure}
        \centerline{\includegraphics[scale=0.8]{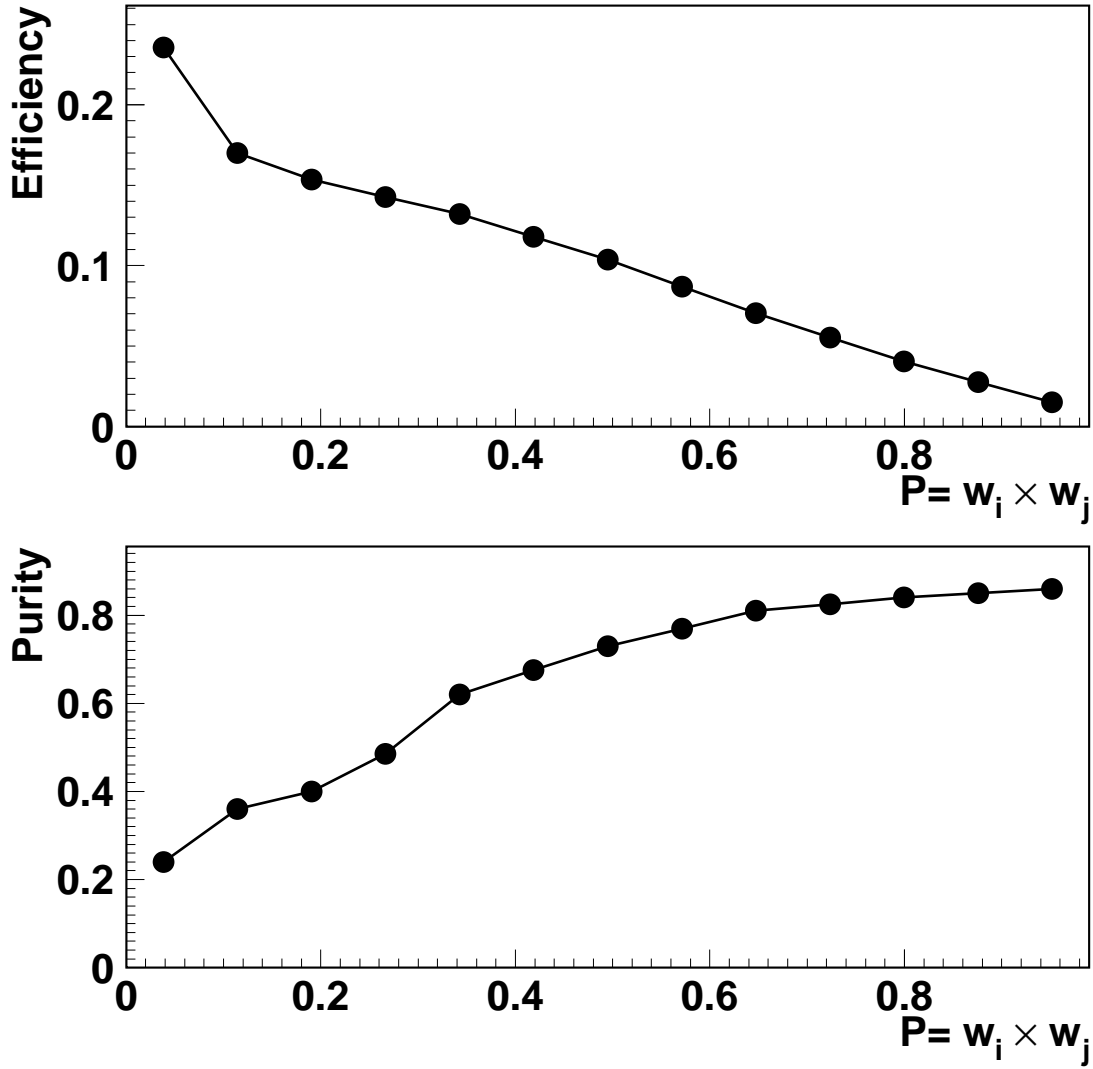}}
        \caption{The $\rm \pi^0 $ reconstruction efficiency (top)
         and the purity (bottom)   
         for different cuts on the weight $ P =w_i\times w_j $ of the 
         $ij$ photon pair.
         The  purity and efficiency are estimated from the JETSET Monte Carlo. 
         The corresponding statistical errors 
         are smaller than 1\%.}
        \label{fig:purity}
\end{figure}

  
  The $ \pi^0 $ reconstruction 
  efficiency and purity are illustrated in Figure~\ref{fig:purity} for different
  cuts on $P.$
  The efficiency is defined as the ratio of the 
  number of correctly reconstructed
   \pins over the number of generated \pins, and 
   the \pin purity is defined as the ratio of signal over 
   total entries in a photon-pair 
   mass window between 100 and 170 MeV.
\section{ Selection of \pin Pairs  }
   The average number of \pins produced in $\rm Z^0$  decays has 
   been measured~\cite{PDGT} to be $ 9.76\pm 0.26,$ which is
   reproduced by our Monte Carlo simulations.
   This leads to about $\rm 45 $ possible \pin pairings per event.
   Considering only \pin candidates with $P>0.1$ 
   (i.e.\ 17\% efficiency and 36\% purity),
   we reconstruct at the detector level 4.7 \pin candidates on average 
   per event. This leads to about $\rm 8 $ pairings among which  
   only 1 pair on average is really formed by true \pins.  
   Here, the detector level 
   means that detector response,
   geometrical acceptance and photon reconstruction efficiency are
   taken into account. Therefore, the \pin pair sample 
   is background dominated and    
   the study of \pin  pair correlations or invariant mass spectra 
   is subject to very large background subtraction.
   Monte Carlo must be used to predict both the shape and amount of background
   to be subtracted, leading to large systematic errors in the measurements
   of the BEC source parameters.\par
\begin{figure}
        \centerline{\includegraphics[scale=0.8]{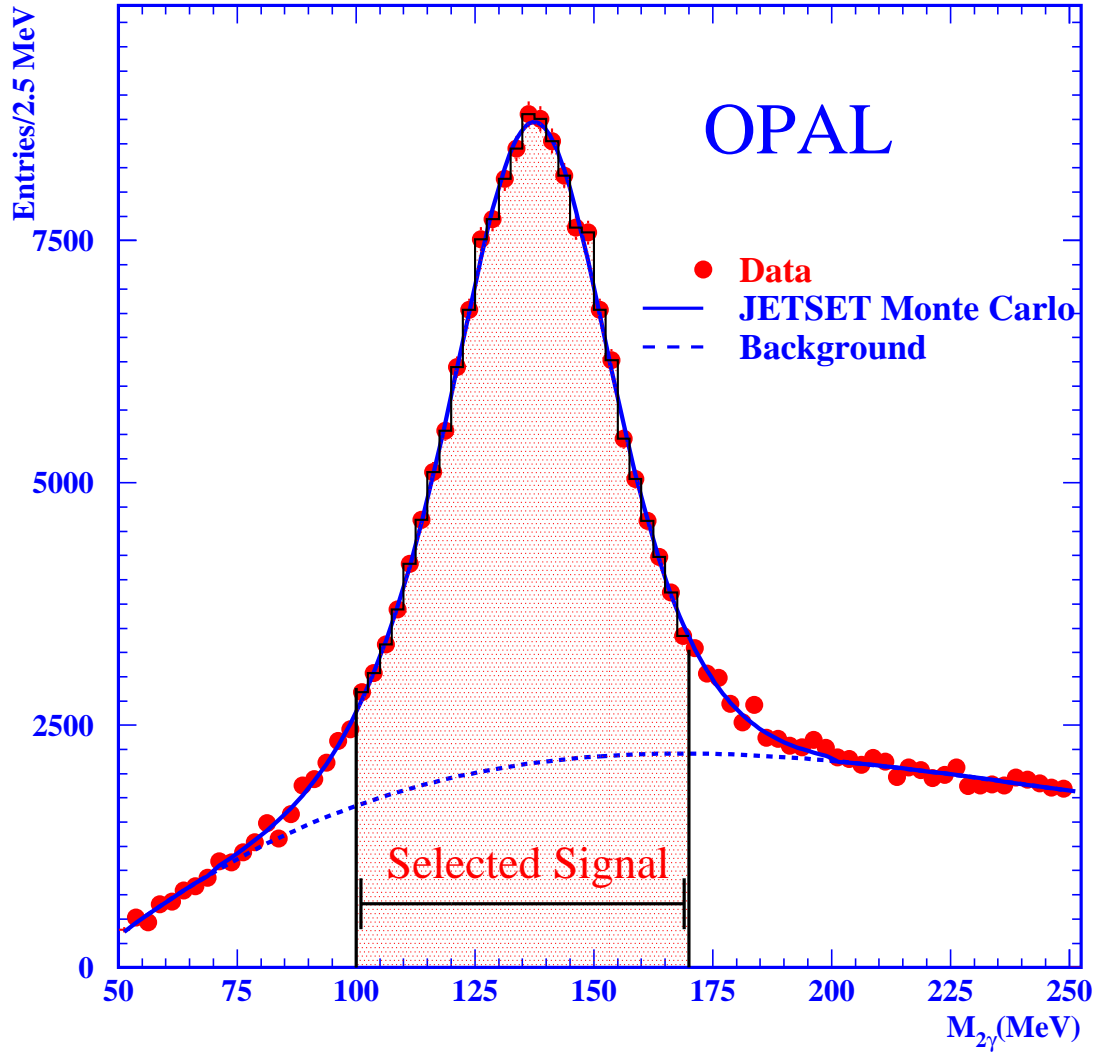}}
        \caption{Distribution of two-photon invariant   
         mass, $M_{2\gamma},$  
         for selected events which have exactly
         two reconstructed \pin candidates per event.
         The smooth curves represent the total Monte 
         Carlo expectation (solid line) and 
         the background (dashed line) expectation.
         The curves are normalised to the same 
         number of total selected hadronic $\rm Z^0 $ decays as in the data.
         The shaded region (100--170 MeV) represents the selected window 
         for the \pin signal.} 
        \label{fig:pi0sel}
\end{figure}

   To avoid this, the \pin  selection criteria are 
   tightened.
   We select \pins which have a momentum above 1 GeV.  This cut
   reduces the fraction of fake \pins. In addition, it removes 
   \pins produced by hadronic interactions in 
   the detector material for which the Monte Carlo simulation
   is not adequate.  The probability $ P $  
   associated to each \pin  candidate is required to 
   be greater than 0.6.  In the case where a photon can be combined in more
   than one pair, only the pair with the highest probability is considered
   as a \pin candidate.
   Among the events with four or more reconstructed photon 
   candidates, only those leading to a possible \pin pair with 
   four distinct photon candidates 
   are retained for further analysis. 
   Events with six or more photon candidates leading to 
   more than two \pin candidates are rejected. 
   They represent about 10\% of the retained sample and 
   would increase the sensitivity to unwanted resonance signals 
   if they were not rejected.
   Figure~\ref{fig:pi0sel} shows the photon pair mass, $M_{2\gamma},$ for
   the selected events. The average 
   purity of the \pin sample is 79\% in the mass window between 
   100 and 170 MeV.
   The background is estimated directly from data by a 
   second order polynomial fit
   to the side bands of the peak and by Monte Carlo simulation. 
   The two background estimations yield compatible 
   results and the Monte Carlo reproduces correctly the data. 
   The superimposed curves are not the result of a fit to the data,   
   but smoothed histograms of 
   the Monte Carlo expectations for signal and background 
   normalised to the total
   number of selected hadronic $\rm Z^0 $ decays.\par
\begin{figure}
        \centerline{\includegraphics[scale=0.8]{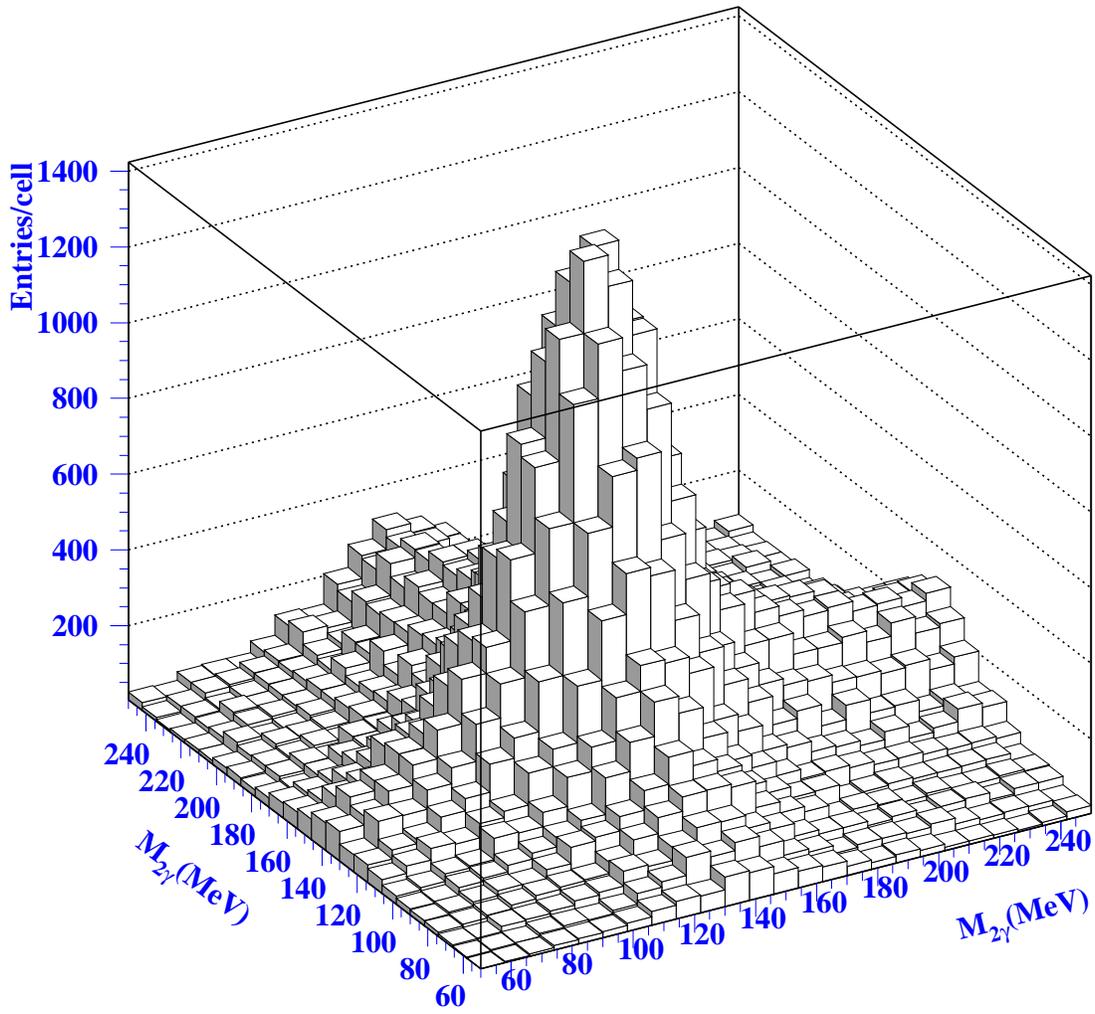}}
        \caption{ The two values of  $ M_{2\gamma} $ for each selected
              event. The cell size is $8\times 8$
          $\rm MeV^2.$ }
        \label{fig:legopi}
\end{figure}

   A clear \pin pair signal is obtained as shown in 
   Figure~\ref{fig:legopi} where the two values of 
   $ M_{2\gamma} $ are shown for the retained events.  
   A \pin pair is considered
   as a signal candidate if both values of $M_{2\gamma}$ 
   are within the mass window between 100 and 170 MeV. 
   The average \pin pair signal purity
   is $\rm 60\% $ and the Monte Carlo simulation
   describes the data well.
   Kinematic fits were made, constraining the mass of pairs of photon 
   candidates to the \pin mass, with the assumption that the photons
   come from the primary interaction vertex. Monte Carlo studies showed that
   this gives a 26\% improvement in the resolution of the 
   \pin momentum.\par 
\section{The BEC Function}
 The correlation function is defined as the ratio,
\begin{equation}  
\label{eq:eq1}
  C(Q) =\frac{\rho(Q)}{\rho_0(Q)}, 
\end{equation}
 where $ Q $  is a Lorentz-invariant variable expressed in terms of
 the two \pin four-momenta 
 $  p_1 $ and  $ p_2$ via $  Q^2= -(p_1-p_2)^2,\ $  $ \rho(Q)=(1/N)
 {\rm d} N/{\rm d}Q $
 is the measured $ Q $ distribution of the two \pins and 
 $ \rho_0(Q) $ is a reference distribution which 
 should, in principle, contain all the correlations included in $ \rho(Q) $ 
 except the BEC.
 For the measurement of $\rho_0(Q), $ we consider the two 
 commonly used methods~\cite{opal0}: 
\begin{itemize}
\item Event Mixing: Mixed \pin
      pairs are formed from \pins belonging to different 
      $\rm Z^0 $ decay events in the data. 
      To remove the ambiguity on how to mix events, we select two-jet
      events having a thrust value $T> 0.9,$
      i.e.\ well defined back-to-back two-jet events.  
      The thrust axes of the two events are required 
      to be in the same direction 
      within ($\rm \Delta\cos{\theta} \times\Delta\phi$)=  
      ($\rm 0.05 \times 10 ^{\circ} $).
      Mixing is then performed 
      by swapping a \pin from one event with a \pin from another 
      event. 
      To avoid detection efficiency problems arising from different
      detector regions, swapping of two pions
      is performed only if they 
      point to the same region of the electromagnetic barrel detector
      within ($\rm \Delta\cos{\theta} \times\Delta\phi$)=  
      ($\rm 0.05 \times 10 ^{\circ} $).  
      With this procedure, we start with two hadronic
      $\rm Z^0$  events each having two \pin candidates and can  
      end up with between zero and four pairs of mixed  
      \pin candidates. The $Q$ variable is then calculated 
      for each of the mixed pairs. 
      If the contributions 
      from background  are removed or suppressed,
      this method offers the advantage of being independent of 
      Monte Carlo simulations, since $ C(Q) $ can be obtained from 
      data alone.
\item Monte Carlo Reference Sample: 
      The $ \rho_0 $ distribution is constructed
      from Monte Carlo simulation without BEC.
      The Monte Carlo is assumed to reproduce correctly all the other 
      correlations present in the data, mainly those corresponding to 
      energy-momentum conservation and those due to known hadron decays.
      In order to be consistent with the first method,
      the cut $ T>0.9 $ is also applied for both data and Monte Carlo.  
\end{itemize}     
   In the following, the distributions $\rho(Q)$ and $\rho_0(Q)$
   are measured from the same sample of selected events. 
   The mixing technique is used as the main analysis
   method and the Monte Carlo reference technique 
   is applied only for comparison.
\section{The Measured BEC Function and Background Contribution}
  The correlation function, $ C(Q),$
  corresponds experimentally to the  
  average number of \pin pairs, corrected for background,
  in the data sample divided by 
  the corresponding  corrected average number in the reference sample.
  Thus, we can write
\begin{equation}
\label{eq:eq3}
 C(Q) = \frac{\rho(Q)}{\rho_0(Q)} = 
      \frac{\rho^{m}(Q)-\rho^{b}(Q)}{\rho_0^{m}(Q)-\rho_0^b(Q)}, 
\end{equation}
 where $\rho^m $ and $\rho^m_0 $ are the measured values, and 
 $\rho^b $ and $\rho^b_0 $ are the corresponding corrections for background 
 contributions. 
 For both the numerator and denominator, the background
 consists mainly of \pin pairs in which one or both $\rm \pi^0 $ candidates 
 are fake.\par 
 The  background distributions $\rho^b $ and $\rho^b_0 $ are  
 obtained from the Monte Carlo information. These background
 distributions can also be obtained 
 from data using a side band fit to the projected spectra of the
 two-dimensional $M_{2\gamma}$ distributions (see Figure~\ref{fig:legopi})
 in each 400 MeV interval of the measured $Q$ variable. 
 The resulting background distributions are correctly reproduced by 
 Monte Carlo. However, for the
 smaller $Q$ intervals as used in this analysis, i.e.\ 100 MeV, 
 the side band fit is subject 
 to large statistical fluctuations, so the Monte Carlo distributions 
 have to be used. \par
 In the region of interest where the BEC effect is observed, $ Q< 700 $ MeV, 
 pion pairs from particle (resonance) decays could mimic the effect. The
 relevant decays are:
   $\rm K_s^0\rightarrow \pi^0\pi^0, $ 
   $\rm f_0(980)\rightarrow \pi^0\pi^0, $ and  
   $\rm \eta\rightarrow \pi^0\pi^0\pi^0 $ with branching ratios of
   39\%, 33\% and 32\% respectively.
   Pion pairs from $\rm \eta $ decay contribute only to the 
   region $Q< 315 $ MeV. 
   According to Monte Carlo studies, 
   the number of reconstructed 
   $\rm K_s^0 $ in the 2 \pin channel is very small.  
   Furthermore, the hypothesis that each \pin 
   originates from the primary 
   vertex, as used in the kinematic fits (Section 4), does not apply.
   This is an advantage for this analysis since the  $\rm K_s^0 $ peak
   is flattened, making its effect on the $Q$ distribution negligible.   
   The Monte Carlo estimates of this particle decay backgrounds are included in 
   the distribution $\rho^b(Q)$, adjusting the rate of individual hadrons to the 
   LEP average~\cite{PDGT} where necessary.\par
\begin{figure}
        \centerline{\includegraphics[scale=0.8]{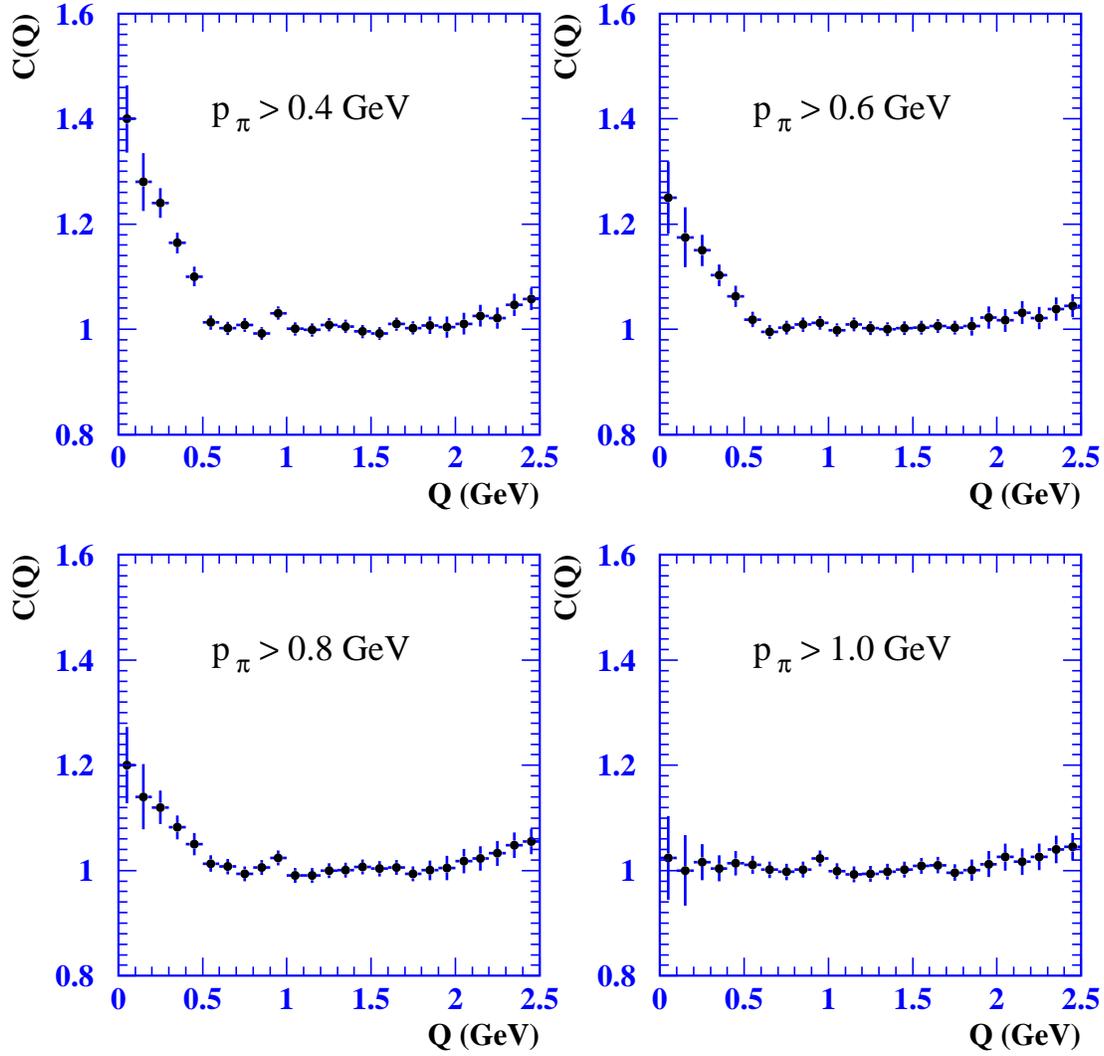}}
        \caption{ The correlation distribution $ C(Q) $ determined
         for JETSET Monte Carlo events (generated without BEC effect) for
         different cuts on the \pin momenta, $p_{\pi}.$
         }     
   \label{fig:genuine} 
\end{figure}

 
   For our analysis we select  
  \pin candidates with momentum greater than 1 GeV. This is  
  dictated by the observation of correlations at small $ Q $ even for 
  Monte Carlo events generated without any BEC effect.
  Indeed, as shown in Figure~\ref{fig:genuine}, a clear BEC-type effect
  is visible in the correlation function obtained from Monte Carlo
  events without BEC for different low cuts on \pin momentum.
  Using Monte Carlo information, we find that these
  correlations are
  mainly caused by \pins originating from secondary interactions with the 
  detector material. They would constitute
  an irreducible background to the BEC effect if low momentum \pins
  are considered in the analysis. 
  This effect vanishes for \pin momenta greater than 1 GeV. \par
  We rely on Monte Carlo simulation only to define the appropriate
  momentum cut (i.e.\ 1 GeV) which completely suppresses
  the effect of soft pions produced in the detector material,  
  rather than relying on its prediction for the exact shape 
  and size of this effect. The reason is that,       
  in contrast to charged pions where the measured track 
  information can be used to 
  suppress products of secondary interactions in the detector material, the
  neutral pions have to be assumed to originate from the main 
  interaction vertex.
  Furthermore, with this assumption the 
  kinematic fits (Section 4) 
  bias the energy of soft
  pions emitted in the detector material towards larger values
  since the real opening angle between the photons is larger
  (vertex closer to the calorimeter) than the assumed one.\par
  With the above selection criteria, the composition of the 
  selected \pin pair sample is studied  using Monte Carlo simulations.
  According to the string fragmentation model implemented
  in JETSET, the selected 
  sample consists of about 97.9\% of mixed pion-pairs from
  different hadron decays, 2\% of pairs  
  belonging to the decay products of the same hadron and only
  0.1\% prompt pairs from the string break-ups. 
  Similarly, using the cluster fragmentation model implemented
  in HERWIG, the selected sample consists of 97\%  of pairs
  from different hadron decays, 2.3\%  
  belonging to the decay products of the same hadron  
  and only 0.7\%  originating
  directly from cluster decays.   
  It is worth mentioning that even if the direct pion pairs 
  from string break-up (JETSET) 
  or cluster decays (HERWIG) 
  were all detected and accepted 
  by the analysis procedure, they would be diluted in  
  combination with other pions and
  would constitute only a marginal 
  fraction ($\rm < 1 \%$ ) of the total number of
  reconstructed \pin pairs. 
  Thus, our analysis has no 
  sensitivity to direct pion pairs originating from string 
  break-up or cluster decay.\par
\section{Results}
\begin{figure}
        \centerline{\includegraphics[scale=0.8]{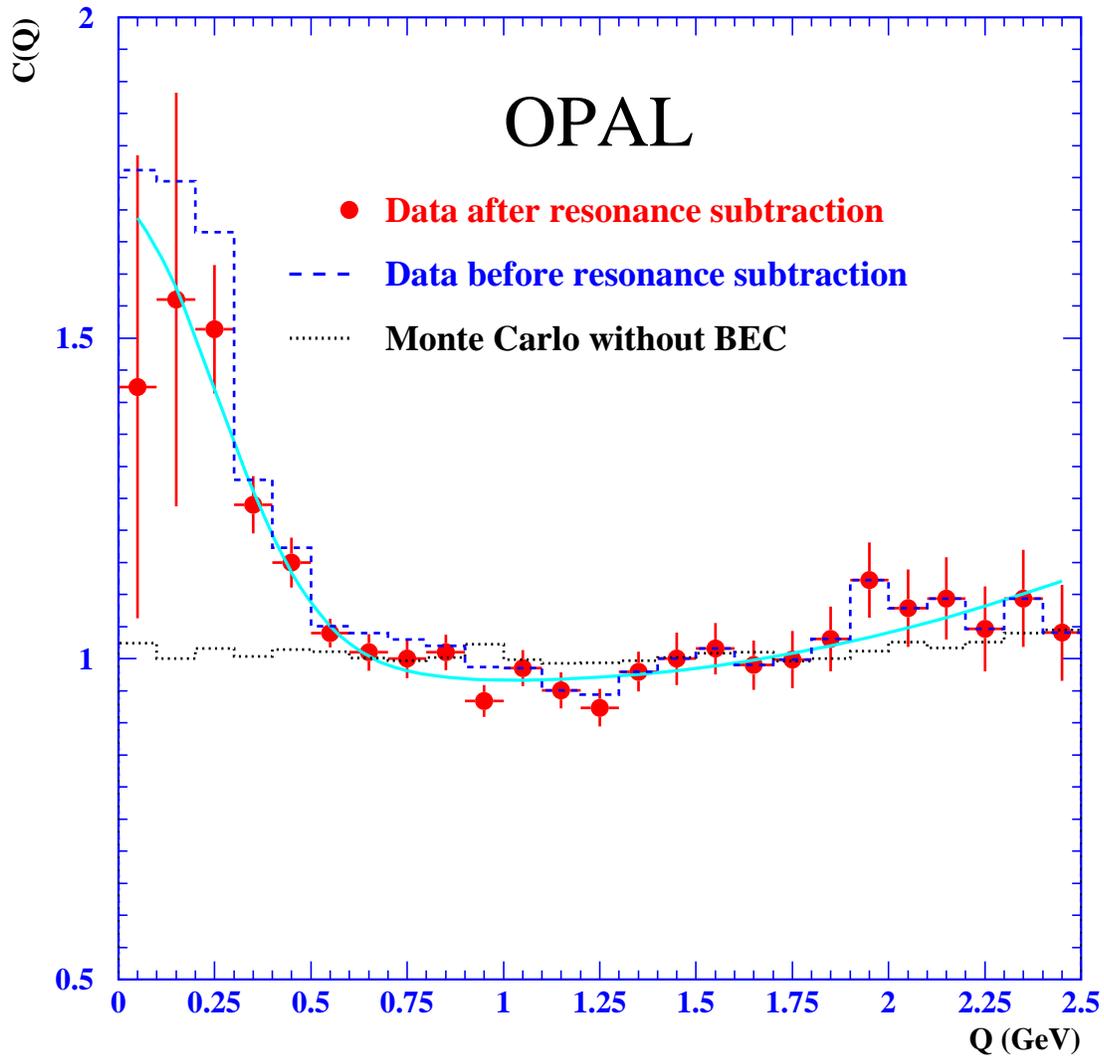}}
        \caption{ The correlation distribution $ C(Q) $ as measured
         for OPAL data. The smooth curve is the fitted correlation function
         and the dotted histogram is the correlation distribution obtained 
         for JETSET Monte Carlo events generated without BEC.
         The dashed histogram represents the measured correlation 
         function before the subtraction of the contributions  
         from known hadron decays.}
        \label{fig:bec} 
\end{figure}

  The correlation distribution $ C(Q)$ (Eq. (\ref{eq:eq3}))
  is parametrised using the Fourier transform of the 
  expression for a static sphere of emitters with 
  a Gaussian density (see e.g.\ \cite{kaons}):
 \begin{equation}
 \label{eq:eq4}
    C(Q) = N(1+\lambda\exp{(-R^2Q^2)})(1 + \delta Q+\epsilon Q^2). 
\end{equation}
  Here $\rm \lambda $ is the chaoticity of the correlation
  [which equals zero for a fully coherent (non-chaotic) source and one
  for a chaotic source], $ R $ is
  the radius of the source, and $ N $ a normalisation factor.
  The empirical term, $(1 + \delta Q+ \epsilon Q^2), $ accounts
  for the behaviour of the correlation function at high $ Q $ due to any 
  remaining long-range correlations. 
  The $C(Q) $ distribution for data is shown in 
  Figure~\ref{fig:bec} as the points with  
  corresponding statistical errors, and the smooth curve is the fitted 
  correlation function in the $Q$ range between 0 and 2.5 GeV.
  A clear BEC enhancement is observed in the low $ Q $ region
  of the distribution.   
  The parameters are determined to be:
$$
\begin{array}{cl}
 \lambda   =& 0.55 \pm 0.10, \\
   R       =& (0.59 \pm 0.08)  \rm\ fm, \\
   N       =& 1.10 \pm 0.08,  \\
   \delta  =&(-0.14 \pm 0.05)\ \rm GeV^{-1}, \\
  \epsilon =&( 0.07\pm  0.03)\ \rm GeV^{-2}, 
\end{array}
$$
 where the quoted errors are statistical only
 and the 
 $ \chi^2/ \rm ndf $ of the fit is 14.7/19.\par
 The distribution $C(Q) $ obtained for Monte Carlo events 
 generated with no BEC is shown as a 
 histogram in the same figure. 
 It shows that there is no residual 
 correlation at low $ Q $  and indicates that
 the observed enhancement is present in the data only.
 The dashed-line histogram of Figure~\ref{fig:bec} represents 
 the correlation function obtained from data
 but before  
 the subtraction, using the Monte Carlo estimates, 
 of pairs from the decay 
 products of the same hadron, indicating that these
 contributions have only a minor influence on the measured parameters.   
 In addition, the correlation function constructed with 
 background \pin pairs does not 
 show any enhancement at low $Q\ $(not shown).
 Here, background \pin pairs are defined as
 pairs for which one or both of the \pins are outside the mass window 100-170 MeV, i.e.\ these are likely to be fake \pin candidates. \par
 The second method, which uses the MC reference sample, 
 yields the following results:
$$\begin{array}{cl}
   \lambda &= 0.50 \pm 0.10, \\
      R    &= (0.46 \pm 0.08) \rm fm .
\end{array} $$  
  These results are quoted for comparison only.
  We choose to quote the results obtained with the event mixing method
  since they are much less dependent on details of the Monte Carlo modelling.\par
  The string model predicts a smaller source radius 
  and a larger chaoticity in the BEC effect for \pin pairs 
  than for $\rm \pi^{\pm} $ pairs, while the cluster model predicts
  no difference. These predictions hold only for prompt boson pairs 
  produced directly from the string or cluster decays. 
  According to our Monte Carlo simulations, we have no sensitivity 
  to these pairs.
\section{ Systematic Uncertainties}
  Potential sources of systematic error are investigated. In each case 
  the effect on the parameters $ R $ and $ \lambda $ and their 
  deviations with respect to the standard analysis are estimated. 
  The results are summarised in Table~\ref{tab:systable}.
 \begin{itemize}
 \item{\bf Bin width resolution:}
     After the kinematic fits (Section 4), the resolution  
     on the invariant mass of two pions, or on the variable $Q,$ is 
     approximately 60 MeV. We have chosen
     a bin width of 100 MeV for the fit to the measured $C(Q)$ distribution. 
     This bin width is varied from 
     100 MeV to 80 MeV and to 120 MeV.
\item{ \bf Fit range:}
     The low end of the fit range is set to start 
     at $ Q = 350 $ MeV (fourth bin)
      The high end of the fit range is changed to stop at  $Q = 2 $ GeV.
\item {\bf Effect of hadron decays:  }
     To estimate the effect of the \pin pairs from the same resonance
     decay on the  measured BEC parameters, the   
     estimated contribution is varied by $\rm \pm 10\% $ which 
     represents the typical error on the measured individual 
     hadron rates~\cite{PDGT}.  
%
  In order to investigate the dependence of the measured parameters $R $ and 
  $\lambda $  on the \pin mo\-men\-tum cut, the analysis is repeated 
  for \pin momenta larger than 1.2 GeV.
\item {\bf Analysis procedure:  }
       The analysis is repeated for several variations 
       of the selection criteria.
  \begin{itemize}
      \item{\bf 1)} The \pin selection mass window is changed from 
            100-170 MeV to 110-165 MeV ( increases the
            \pin purity by 5\%).
      \item{\bf 2)} The probability   
             for \pin selection is changed from 
            0.6 to 0.5 (reduces the \pin purity by 5\%).
      \item{\bf 3)} The thrust value for two-jet events is changed from 0.9 to 
            0.85 and to 0.92 (changes the overall event sample
            size by $\pm 5\%$).
      \item{\bf 4)} The factor  $ 1+\delta Q + \epsilon Q^2 $ is replaced 
             by $ 1+\delta Q. $ 
      \item{\bf 5)} \pin from different events are mixed if they point to the same 
            region of the detector within  
        $( \Delta\cos{\theta} \times\Delta\phi) = 
            (0.10 \times 15 ^{\circ} )$ instead of   
            $( 0.05 \times 10 ^{\circ}).$ 
   \end{itemize}   
 \end{itemize}
   The final systematic errors are obtained from quadratically adding 
   the deviations from the central value. Thus, 
$$\begin{array}{cl}
   \lambda &= 0.55 \pm 0.10 \pm 0.10, \\
     R    &= (0.59 \pm 0.08 \pm 0.05)\ \rm fm.
\end{array} $$ \par
 \section{Conclusions}
  We have observed Bose-Einstein correlations of \pin pairs produced
  in hadronic $\rm Z^0 $ decays. Assuming a Gaussian shape for the source, we
  obtain $ \lambda = 0.55 \pm 0.10 \pm 0.10 $ for the chaoticity parameter 
  and $R = (0.59 \pm 0.08 \pm 0.05)\ \rm fm $  for the radius.
  In order to construct a reference sample with the event mixing method, 
  this analysis is restricted 
  to well defined back-to-back two-jet events. Furthermore, in order
  to remove \pins not originating from the primary interaction vertex
  the considered momentum
  phase space is restricted to $p_{\pi^0}>1 $ GeV.
  The measured value of the source radius is smaller than     
  our former value~\cite{opallast},
  $ R=(1.002 \pm 0.016 ^{+0.023}_{-0.096})\ \rm fm, $ obtained for 
  charged pions for which the measured track parameters allowed
  access to lower momenta and
  where the reference sample was 
  constructed with unlike-sign pion pairs.     
  However, the value is  
  compatible with the 
  LEP inclusive average~\cite{Alexander}, 
  $R=(0.74\pm0.01\pm 0.14)\ \rm fm$,
  for charged pions.
  Pions from strong decays constitute the dominant part of our sample of
  reconstructed \pin pairs.  
  We have no sensitivity to test the 
  string or cluster model predictions concerning differences 
  between neutral and
  charged pion pairs. We deduce that Bose-Einstein correlations
  exist between $\pi^0$ pairs in which each \pin is
  a strong decay product of a different hadron.
\par 
\begin{table}
\caption{\large Systematic errors}
   \begin{center}
       \begin{tabular}{|l|c|c|c|c|}
       \hline
       $ Item $ & $ \lambda\ $  &  $R$ [fm]  & $\Delta \lambda$  &$\Delta R $  \\
       \hline
       \hline
Basic result &
$\rm 0.55 \pm 0.10 $& $\rm 0.59\pm 0.08 $  &+0.00 & +0.00  \\
       \hline      
Bin width=80 MeV&
$\rm 0.54 \pm 0.13 $& $ 0.58 \pm 0.12 $   &-0.01 & -0.01   \\
       \hline      
Bin width=120 MeV&
$ 0.57 \pm 0.09    $&   $ 0.60 \pm 0.07 $ &+0.02 & +0.01  \\
       \hline
Low end of the Fit range = 350 MeV&
    $ 0.64 \pm 0.14 $&   $ 0.62 \pm 0.12 $ &+0.09 & +0.03  \\
      \hline      
High end of the Fit range = 2 GeV&
    $ 0.58 \pm 0.11 $&   $ 0.56 \pm 0.10 $ &+0.03 & -0.03  \\
      \hline      
Resonance contribution +10\% &
$\rm 0.54 \pm 0.10 $& $\rm 0.59\pm 0.08 $&-0.01 & +0.00   \\
       \hline
Resonance contribution -10\% &
$\rm 0.55 \pm 0.10 $& $\rm 0.58\pm 0.09 $&-0.00 & -0.01   \\
       \hline
Momentum cut = 1.2 GeV &
$\rm 0.53 \pm 0.11 $& $\rm 0.60\pm 0.08 $&-0.02 & +0.01   \\
       \hline
Analysis procedure: &   &  &   &                           \\
  1) $\pi^0-$signal mass window  & $\rm 0.55 \pm 0.10\ $&$\rm 0.58\pm 0.09\ $& +0.00 & -0.01\\
  2) Photon-pair probability  & $\rm 0.57 \pm 0.09\ $&$\rm 0.57\pm 0.08\ $& +0.02 & -0.02\\
  3) Thrust value   & $\rm 0.55 \pm 0.10\ $&$\rm 0.60\pm 0.09\ $& +0.00 & +0.01\\
  4) Long range corr. term & $\rm 0.56 \pm 0.10\ $&$\rm 0.58\pm 0.10\ $& +0.01 & -0.01\\
  5) Mixing condition  & $\rm 0.54 \pm 0.08\ $&$\rm 0.59\pm 0.07\ $& -0.01 & +0.00\\
       \hline       
       \hline       
Total sys. error &  &  & 0.10 & 0.05 \\
\hline
       \end{tabular}\label{tab:systable}
     \end{center}
\end{table}
\medskip
\bigskip\bigskip\bigskip
\appendix
\par
\section*{Acknowledgements} 
We particularly wish to thank the SL Division for the efficient operation
of the LEP accelerator at all energies
 and for their close cooperation with
our experimental group.  In addition to the support staff at our own
institutions we are pleased to acknowledge the  \\
Department of Energy, USA, \\
National Science Foundation, USA, \\
Particle Physics and Astronomy Research Council, UK, \\
Natural Sciences and Engineering Research Council, Canada, \\
Israel Science Foundation, administered by the Israel
Academy of Science and Humanities, \\
Benoziyo Center for High Energy Physics,\\
Japanese Ministry of Education, Culture, Sports, Science and
Technology (MEXT) and a grant under the MEXT International
Science Research Program,\\
Japanese Society for the Promotion of Science (JSPS),\\
German Israeli Bi-national Science Foundation (GIF), \\
Bundesministerium f\"ur Bildung und Forschung, Germany, \\
National Research Council of Canada, \\
Hungarian Foundation for Scientific Research, OTKA T-029328, 
and T-038240,\\
The NWO/NATO Fund for Scientific Reasearch, the Netherlands.\\

\end{document}